# Randomization Adaptive Self-Stabilization

Shlomi Dolev *    Nir Tzachar †

August 14, 2018

## Abstract

We present a scheme to convert self-stabilizing algorithms that use randomization during and following convergence to self-stabilizing algorithms that use randomization only during convergence. We thus reduce the number of random bits from an infinite number to a bounded number. The scheme is applicable to the cases in which there exits a local predicate for each node, such that global consistency is implied by the union of the local predicates. We demonstrate our scheme over the token circulation algorithm of Herman [10] and the recent constant time Byzantine self-stabilizing clock synchronization algorithm by Ben-Or, Dolev and Hoch [3]. The application of our scheme results in the first constant time Byzantine self-stabilizing clock synchronization algorithm that uses a bounded number of random bits.

## 1 Introduction

Self-stabilizing algorithms are designed to start from an arbitrary state and eventually exhibit a desired behaviour. Thus, a self-stabilizing algorithm can recover once the assumptions made for its convergence from the arbitrary state hold. Self-stabilizing algorithms that use randomization are able to achieve tasks that cannot be achieved by deterministic means. In some cases, randomization allows faster convergence of self-stabilizing algorithms. Often, randomized self-stabilizing algorithms are designed to use an infinite amount of random bits to operate correctly. However, the creation of (real) random bits is considered expensive; thus, a *randomization adaptive* self-stabilizing algorithm that uses random bits during convergence but does not use random bits following the convergence is desirable. Such a notion of adaptiveness has been studied in the past, where the resource demands of a self-stabilizing algorithm are reduced upon convergence, be it memory requirements, e.g., [2] or communication requirements, e.g., [5].

*Department of Computer Science, Ben-Gurion University of the Negev, Beer-Sheva, 84105, Israel, dolev@cs.bgu.ac.il. Partially supported by Deutsche Telekom, Rita Altura Trust Chair in Computer Sciences and the ICT Programme of the European Union under contract number FP7-215270 (FRONTS).

†Department of Computer Science, Ben-Gurion University of the Negev, Beer-Sheva, 84105, Israel, tzachar@cs.bgu.ac.il. Partially supported by Deutsche Telekom and the Lynne and William Frankel Center for Computer Sciences and the ICT Programme of the European Union under contract number FP7-215270 (FRONTS).

We suggest a new scheme for converting randomized self-stabilizing algorithms to randomization adaptive self-stabilizing algorithms. Our scheme can be employed when every execution of the algorithm, starting in a safe configuration, is legal, regardless of the random input, as opposed to legal with some probability. Our scheme can also be applied in case there is at least one random sequence that implies liveness from every safe configuration. Generally speaking, the scheme is based on collecting the entire history of the system at each node, and examining this history to check if the algorithm has converged. If so, no randomization is used. We demonstrate the scheme on a leader election algorithm derived from the token circulation algorithm of [10]. We also obtain token circulation with deterministic behaviour following convergence.

When Byzantine nodes are introduced, we suggest a scheme based on *unreliable detectors*. The unreliable detector gives an unreliable indication whether the algorithm is in a safe configuration. Roughly, when the algorithm has converged, the detector *eventually* notifies all nodes on this fact; before the algorithm converges the detector notifies at least one node that randomization is still needed. The other part of the conversion scheme is a *randomization surrogate*, where each node that gets an indication on non-convergence from the unreliable detector supplies random bits to all the nodes including itself; each node xors the random bits receive from all nodes in the system and uses the result as its source for random bits. In the presence of Byzantine nodes, such randomization assistance must be done atomically, avoiding Byzantine nodes choice' that can potentially nullify the random bits selected.

We demonstrate our second scheme over the non-randomization adaptive self-stabilizing clock synchronization algorithm presented in [3]. The application of our scheme results in the first constant time Byzantine self-stabilizing clock synchronization algorithm that uses a bounded number of random bits.

Similar goals to ours have been investigated, in particular by Rao [9], where the author seeks *eventual determinism* which is similar in nature to the concept of randomization adaptiveness. Rao also describes two particular applications, providing eventual deterministic algorithms to the symmetric dining philosophers problem (based on a combination of the probabilistic algorithm of Lehmann and Rabin [8] and the deterministic algorithm of Chandy and Misra [8]) and to the token circulation algorithm of Herman [10] – which is more memory efficient than the one we present. However, Rao does not present a generic scheme to convert randomized algorithm to randomization adaptive ones, nor does Rao present a solution which can work in the presence of Byzantine nodes. In contrast, we provide both a generic scheme for the general case and for the Byzantine case.

**Paper organization.** The next section describes particular synchronous settings for distributed systems; the particular settings are chosen to simplify our presentation. In Section 3 we describe our conversion scheme where there are no Byzantine nodes, and in Section 4 we describe the modifications needed when Byzantine nodes are introduced.

## 2 The System

The *system* is modeled as a set of *nodes*, $N = \{p_1, p_2, \ldots, p_n\}$, each node is a state machine. A node may communicate with other nodes by sending *messages*. The communication paths between nodes are modeled as an undirected graph, $G = (V, E)$, where nodes are the vertices of $V$, and two nodes, $u$ and $v$, may communicate if $(u, v) \in E$. When dealing with Byzantine nodes, we assume the communication graph is complete.

For the sake of discussion simplicity, we assume that the system is synchronous, meaning that, repeatedly, a shared *pulse* arrives at all nodes simultaneously. The time duration between two successive pulses is called a *round*. At the beginning of each round, each node, $p$, sends messages to all other nodes and receives all messages sent to $p$ before the next pulse.

A *state*, $s_p^i$, of a node $p$ is the memory contents of $p$ at the beginning of round $i$. A configuration, $c^i$, representing the state of the system at the beginning of round $i$, is defined by the set of states of all nodes at the beginning of round $i$: $c^i = \{s_{p_1}^i, s_{p_2}^i, \ldots, s_{p_n}^i\}$.

An *execution* of the system, $\mathcal{E}$, is a sequence of infinite system configurations, $\{c^i\}_{i=0}^{\infty}$, such that for each pair of configurations, $c^i, c^{i+1}$, each node $p$ makes the transition from $c_p^i$ to $c_p^{i+1}$ according to $p$'s algorithm, state and messages received by $p$ during the $i$'th round.

A *task* is defined by a set of executions called *legal executions* and denoted $\mathcal{LE}$. A configuration $c$ is a *safe configuration* for a system and a task $\mathcal{LE}$ if every execution that starts in $c$ is in $\mathcal{LE}$. A system is *self-stabilizing* for a task $\mathcal{LE}$ if every infinite execution reaches a safe configuration with relation to $\mathcal{LE}$, regardless of the initial configuration. We sometimes use the term "the algorithm stabilizes" to note that the algorithm has reached a safe configuration with regards to the legal execution of the corresponding task. We say that an algorithm is a *randomized* self-stabilizing algorithm if the algorithm makes use of random bits, such that every random input has a positive probability to be used. Moreover, we require that once the execution of a randomized self-stabilizing algorithm has reached a safe configuration, the execution remains in $\mathcal{LE}$ (as opposed to remain in $\mathcal{LE}$ with some probability). A *randomization adaptive* self-stabilizing algorithm is a randomized self-stabilizing algorithm such that in every execution of the algorithm there exists a deterministic infinite suffix of the execution (where no node uses random inputs).

A node, $b$, is a *Byzantine* node if it may act in an arbitrary way. A Byzantine node may not follow the algorithm and may send arbitrary messages to other nodes. We limit the number of Byzantine nodes by $f$, such that $n \geq 3f + 1$. We do not assume that nodes can communicate over private links; our scheme remains correct even when Byzantine nodes can eavesdrop on all messages sent in the system.

## 3 The Conversion Scheme, without Byzantine Nodes

The general conversion scheme we present in this section can be applied to any *adaptable* randomized self-stabilizing algorithm $\mathcal{A}$;

**Definition 3.1.** A randomized self-stabilizing algorithm $\mathcal{A}$ for a task $\mathcal{LE}$ is *adaptable* if, once the system is in a safe configuration, every suffix of an execution that follows is legal (as opposed to legal with some probability). Moreover, for each node, $q$, there exists a self-stabilizing algorithm $\mathcal{P}_q$ – the detector – such that, after the detector has stabilized,
- if the system is not in a safe configuration with regards to $\mathcal{LE}$, all detectors return *false*.
- each execution has an infinite suffix in which no detector returns *false*.

Next, we show that given an adaptable randomized self-stabilizing algorithm $\mathcal{A}$ for a task $\mathcal{LE}$, our scheme produces a randomization adaptive self-stabilizing algorithm $\mathcal{A}'$ which uses a bounded number of random bits. We start with the definition of $\mathcal{A}'$:

**Definition 3.2.** Given an adaptable randomized self-stabilizing algorithm $\mathcal{A}$, define $\mathcal{A}'$ to be the following algorithm: at each round, each node $q$ initiates a new instance of the random bits production protocol (Figure 1). Moreover, at each round, each node $q$ uses the output of the most recently terminated random bit production algorithm as $q$'s random input to $\mathcal{A}$.

```
upon Pulse:
1    initiate a new instance of P_i
2    if the last terminated P_i = false then
3    |    rand_i ← Random()
4    else
5    |    rand_i ← 1 //or a vector of 1's
6    fi
```

Figure 1: Random Bits Production Protocol for Node $i$

The next Lemma states that after an adaptable randomized self-stabilizing algorithm has converged to a legal execution, the random input is no longer necessary:

**Lemma 3.1.** *Let $\mathcal{A}$ be an adaptable randomized self stabilizing algorithm for a task $\mathcal{LE}$. Let $\mathcal{E} = \left\{c^i\right\}_{i=0}^{\infty}$ be an execution, where $c^0$ is a safe configuration. Let $r_i$ be the sequence of random bits used by each node $p_i$. $\mathcal{E} \in \mathcal{LE}$, regardless of the values used for each $r_i$.*

*Proof.* The proof follows from the definition of adaptable randomized self-stabilization. As the execution starts from a safe configuration, $c^0$, any input $r_i$ given to node $p_i$ at round 0 has a positive probability to be drawn randomly. Since the execution must stay in $\mathcal{LE}$ when $r_i$ was truly a random value, $c^1$ must also be a safe configuration. Using an inductive argument, any configuration that follows $c^1$, regardless of the random choices made, is a safe configuration, and hence $\mathcal{E} \in \mathcal{LE}$. □

Next, we establish that $\mathcal{A}'$ is a randomization adaptive self-stabilizing algorithm for the same task as $\mathcal{A}$, $\mathcal{LE}$.

**Lemma 3.2.** *Let $\mathcal{E}$ be an execution of $\mathcal{A}'$, which is legal with regards to the detection algorithm. In each configuration $c^i \in \mathcal{E}$, such that $c^i$ is not a safe configuration with regards to $\mathcal{LE}$, all nodes in the system use random bits as input to $\mathcal{A}$. Moreover, there exists an infinite suffix of $\mathcal{E}$ in which no randomization is used.*

*Proof.* From Definition 3.1, for each node $q$, $\mathcal{P}_q$ returns *false* in round $i$, when $c^i$ is not a safe configuration. On the other hand, also from Definition 3.1, there exists an infinite suffix of $\mathcal{E}$, $\mathcal{S}$, in which all detectors return *true*. Hence, no randomization is used during $\mathcal{S}$. □

Combining Lemmas 3.1 and 3.2, and the self-stabilizing composition scheme from [4] (with regards to the self-stabilizing detector), we get the following theorem:

**Theorem 3.1.** *Given a adaptable randomized self stabilizing algorithm $\mathcal{A}$ for a task $\mathcal{LE}$, $\mathcal{A}'$ is a randomization adaptive self-stabilizing algorithm, which uses a bounded number of random bits, for the same task $\mathcal{LE}$.*

### 3.1 A Generic Detector

Here we present a generic self-stabilizing detector algorithm, which works correctly as long as no Byzantine nodes are present in the system. The detector is based on collecting, at each node, a historical view of the system, such that at each round each node has the correct state of the system $d$ rounds in the past, where $d$ is the diameter of the communication graph. The technique to achieve such history collection appears in Figure 2, which is based on the technique presented in [1]. Next, at each round $i$, each node $q$ examines the history of round $i - d$ to check if the system was in a safe configuration. If the system was in a safe configuration in round $i - d$, $q$ uses no random bits. Otherwise, $q$ uses random bits.

```
variables:
N = the neighbors of i
history_i = [c_{-1}, c_{-2}, ..., c_{-d}]
s_i = the current state of node i

upon Pulse:
1    history_i = [{s_i}, c_{-1}, c_{-2}, ..., c_{-d+1}]
2    for each p ∈ N do
3    |   Send history_i to p
4    done
5    for each history_p Received do
6    |   merge history_i and history_p
7    done

HistDetec()
8    Return c_{-d} is safe
```

Figure 2: History Collection Algorithm for Node $i$

The history collection algorithm presented in Figure 2 maintains, at each node, an array of the last $d$ system configurations – the $history_i$ array. At each pulse, each node $i$ first shifts $i$'s $history_i$ array to the right, discarding $c_{-d}$, and placing the partial configuration $\{s_i\}$ as in $history_i[0]$ (line 1). $i$ then proceeds to send $history_i$ to all of $i$'s neighbors in lines 2-4. Finally, $i$ merges the histories $i$ receives from $i$'s neighbors with $i$'s history array.

**Lemma 3.3.** *The algorithm in Figure 2 is a self-stabilizing algorithm for the following task; at the end of each round $i$, each node holds the same, correct, state of the system at round $i - d$. Moreover, the stabilization time of the algorithm is exactly $d$ rounds.*

Next, we show that the detector, **HistDetec**, which appears in Figure 2 line 8, indeed satisfies Definition 3.1. In the following two Lemmas, we make the following assumptions; let $\mathcal{A}$ be a randomized self-stabilizing algorithm for a task $\mathcal{LE}$, such that, in every execution $\mathcal{E}$, once the system is in a safe configuration every suffix of $\mathcal{E}$ that follows is legal (as opposed to legal with probability). Let all nodes participate in the history collection algorithm, and assume the history collection algorithm has stabilized.

**Lemma 3.4.** *If the system is not in a safe configuration with regards to $\mathcal{LE}$, all **HistDetec** invocations in all nodes return false.*

*Proof.* The key observation is that once the system enters a safe configuration, the execution will remain in $\mathcal{LE}$. Hence, if the system, at round $i$, is not in a safe configuration, it immediately follows that at round $i - d$ the system was not in a safe configuration. As the history collection algorithm has already stabilized, each node holds the correct configuration of the system at round $i - d$, and the **HistDetec** procedure will return *false* at all nodes. □

**Lemma 3.5.** *Let $\mathcal{S} = \{c^i, c^{i+1}, \ldots\}$ be a suffix of $\mathcal{E}$ which is in $\mathcal{LE}$. Then, for all $k \geq d$, all **HistDetec** invocations in all nodes return true.*

*Proof.* As the history collection algorithm has stabilized, for each $k \geq d$, at round $i + k$ all nodes will have the correct configuration of the system at round $i + k - d$. Since $i + k - d \geq i$, it follows that $c_{i+k-d} \in \mathcal{S}$, which implies that **HistDetec** will return *true* at all nodes. □

Combining both of the Lemmas above, we draw the following Theorem, which provides the generic conversion scheme:

**Theorem 3.2.** *Let $\mathcal{A}$ be a randomized self-stabilizing algorithm for a task $\mathcal{LE}$, such that, in every execution $\mathcal{E}$, once the system is in a safe configuration every suffix of $\mathcal{E}$ that follows is legal (as opposed to legal with probability). Combining $\mathcal{A}$ with the history collection algorithm and the detector based on it results in an adaptable randomized self-stabilizing algorithm which obeys Definition 3.1.*

We note that in asynchronous settings, a self-stabilizing snapshot may serve as the detector [11]. However, whenever nodes are anonymous, different techniques are required. For example, when each node knows the topology of the network, the history collection algorithm can still be employed.

Another aspect to consider is a possible weakening of Definition 3.1. Instead of requiring that as long as the algorithm has not stabilized, all detectors return *false*, we may only require that at least one detector returns *false*. In such a case, at each round, each node $i$, such that $i$'s detector returns *false*, will draw enough random bits for the entire system and store them as part of $i$'s history. We then use the same history collection algorithm as above to collect the system history at all of the nodes. Whenever a node $p$ needs to use random bits, $p$ xors the random bits generated for $p$ by all of the nodes $d$ rounds in the past, using the history collected at $p$. A similar argument to Lemma 3.4 can be used to show that the conversion scheme is correct.

### 3.2 Case Study – Token Circulation on an Odd Ring

Herman [10] presents a randomized self-stabilizing algorithm for token circulation on an odd ring. Each node maintains a single bit, holding either 1 or 0. The ring is oriented, and when a node has the same bit as its left neighbor, the node is said to hold a token. The algorithm ensures that, eventually, only one token exists in the system. Moreover, this single token may circulate in the ring to adjacent nodes. The algorithm maintains that, at each round, each node $i$ sends $i$'s bit to $i$'s right neighbor. Each node $i$ then changes $i$'s state according to the following rule, given that $b_i$ holds $i$'s bit, and $b_l$ holds $i$'s left neighbor's bit:

$$b_i = \begin{cases} b_l & \text{if } b_i \neq b_l \\ random & \text{if } b_i = b_l \end{cases}$$

The token circulation algorithm can be adapted to a randomized self-stabilizing leader election algorithm, such that eventually there exists a single (possibly moving) leader in the system. The node holding the token is regarded as the leader.

To apply the generic scheme of the previous section, we first formally define the token circulation task, $\mathcal{TC}$, which suffices for the leader election task. Next, we show that for each execution of the token circulation algorithm, there exists a suffix which is legal (in contrast to legal with probability). It then easily follows that the token circulation algorithm also solves the leader election task and, furthermore, the generic conversion scheme is applicable in this case.

**Definition 3.3.** Given a synchronous ring of nodes, the token circulation task, $\mathcal{TC}$, is defined as all executions in which there exists exactly one token, such that in each $\mathcal{E} \in \mathcal{TC}$, between two consecutive configurations $c_i$ and $c_{i+1}$ in $\mathcal{E}$, the token either stays in the same node or moves one node counterclockwise.

In [10] Herman shows that the algorithm is indeed self-stabilizing with regard to $\mathcal{TC}$, and further establishes that each execution starting in a safe configuration with regards to $\mathcal{TC}$ stay in $\mathcal{TC}$, at each round. It follows that the token circulation algorithm is an adaptable randomized self-stabilizing algorithm for the task $\mathcal{TC}$, which, using our generic conversion algorithm, can be converted to a randomization adaptive self-stabilizing algorithm for the (possibly moving) leader election task.

**Theorem 3.3.** *There exists a randomization adaptive self-stabilizing algorithm for the (possibly moving) leader election task on an odd ring.*

We wish to consider two more aspects of token circulation algorithm: history aggregation and liveness. As can be observed in the token circulation algorithm, the history collection algorithm may not always need to collect the entire history of the system, but may use some form of aggregation. For example, each node needs only remember if the number of tokens in the system is 0, 1 or greater than 1. Combining two disjoint views of the system is trivial, and the resulting history collection algorithm requires memory which is linear with the diameter of the ring.

Finally, in [10], the *liveness* of the system is ensured by the use of randomization; with probability 1, the token circulates the ring ad-infinitum. In other words, our generic scheme cannot be applied in this case. However, upon closer examination, once the system has stabilized the token can be circulated in a deterministic fashion; we can then augment our detector such that, after stabilization is detected, the random input generated by our conversion scheme will ensure the token is passed each round. We can also extend this idea to the general case; if there exists an input for the algorithm, such that starting from any safe configuration and using this input as the randomness for the algorithm the system remains legal, using this input in the detector results in a randomization adaptive algorithm.

## 4 The Conversion Scheme, with Byzantine Nodes

We next present a slight variation of the conversion scheme, designed to operate in the presence of Byzantine nodes – as Byzantine nodes may corrupt the history viewed by nodes. We assume that the communication graph is complete, to prevent Byzantine nodes from interfering with communications between non-Byzantine nodes.

Our conversion scheme can be applied to any *Byzantine adaptable* randomized self-stabilizing algorithm, $\mathcal{A}$ as we now define. Changes from the definition of adaptable randomized self-stabilizing algorithms are emphasized.

**Definition 4.1.** A randomized self-stabilizing algorithm is denoted *Byzantine adaptable* if, once the system is in a safe configuration, every suffix of an execution that follows is legal (as opposed to legal with some probability). Moreover, for each *(non-Byzantine)* node $q$, there exists an algorithm $\mathcal{P}_q$ – the detector – such that:
- When the system is not in a safe configuration, *at least one detector* returns *false*.
- Each execution has an infinite suffix in which all detectors always return *true*.

We use the algorithm in Figure 3 in a *pipelined* fashion, at each node, to generate the required amount of random bits in each round: at each round we start a new instance of $\mathcal{P}_i$ and supply the random bits generated by the most recently terminated instance to the self-stabilizing algorithm. The underlying idea is to use randomness surrogates, such that as long as one correct node observes that the system is not in a safe configuration, this node will generate randomness for all other correct nodes.

```
input: 𝒫_i, the detector
1   if 𝒫_i = false then
2       for each node j do
3           rand ← Random()
4           Send(rand) to j
5       done
6   else
7       for each node j do
8           Send(⊥) to j
9       done
10  fi
11  let rand_j ← Receive from j
12  rand_i ← ⊕_j rand_j
13  set r_i ← rand_i as the random input
```

Figure 3: Random Bits Production Protocol for Node $i$

Next, we show that given a Byzantine adaptable randomized self stabilizing algorithm $\mathcal{A}$ for a task $\mathcal{LE}$, our scheme produces a randomization adaptive self-stabilizing algorithm $\mathcal{A}^b$ which uses a bounded number of random bits. We start with the definition of $\mathcal{A}^b$:

**Definition 4.2.** Given a Byzantine adaptable randomized self-stabilizing algorithm $\mathcal{A}$, define $\mathcal{A}^b$ to be the following algorithm: at each round, each node $q$ initiates a new instance of the random bits production protocol (Figure 3). Moreover, at each round, each node $q$ uses the output of the most recently terminated random bit production algorithm as $q$'s random input to $\mathcal{A}$.

We now turn our attention to show that $\mathcal{A}^b$ is indeed a randomized self-stabilizing algorithm, for the same task $\mathcal{LE}$. We first show that as long as the execution is not in $\mathcal{LE}$, each node will use random bits as input to $\mathcal{A}$, supplied by at least one randomness surrogate.

**Lemma 4.1.** *Let $\mathcal{E}$ be an execution of $\mathcal{A}^b$. In each configuration $c^i \in \mathcal{E}$, such that $c^i$ is not a safe configuration, all nodes in the system use random bits as input to $\mathcal{A}$.*

*Proof.* From Definition 4.1, for at least one node, $q$, $\mathcal{P}_q$ returns *false* in round $i$. Each node, $p$, will then receive random bits from $q$ (including $q$ itself). $p$ will then xor these bits with all the bits $p$ received from other nodes, resulting in a random string of bits. It is also straightforward to observe that any non-random bits sent by nodes for which the indicator returned *true* or from Byzantine nodes, may not harm the randomness of the string generated at $p$. □

Combining Lemmas 4.1 and 3.2, we get the following theorem:

**Theorem 4.1.** *Given a adaptable randomized self stabilizing algorithm $\mathcal{A}$ for a task $\mathcal{LE}$, $\mathcal{A}^b$ is a randomization adaptive self-stabilizing algorithm, which uses a bounded number of random bits, for the same task $\mathcal{LE}$.*

## 4.1 Case Study – Byzantine Clock Synchronization Algorithm

As a case study for applying our technique, we examine the fast randomized self-stabilizing algorithm for Byzantine clock synchronization [3] that, unfortunately, unlike [6], requires randomization following convergence [1]. The algorithm first reduces the clock synchronization problem to Byzantine agreement – if all nodes can agree on a common clock value, the algorithm easily stabilizes. The Byzantine agreement protocol employs a shared coin protocol, which ensures that if all nodes toss a coin each round, then with a constant probability (albeit small), all nodes produce the same random coin.

The clock synchronization algorithm of [3] is composed of two complementing parts: a 2-clock (a clock with two values) synchronization algorithm (which is easily extended to a 4-clock) and a $k$-clock synchronization algorithm which builds on the 4-clock, for an arbitrary $k$. We focus only on the first part of the scheme, the 2-clock synchronization algorithm. It is rather straightforward to apply our techniques to the $k$-clock synchronization algorithm, by using the detector presented in Figure 4. The key observation needed is that, once the $k$-clock synchronization algorithm has converged, the execution remains legal regardless of the random bits used. It follows that the $k$-clock synchronization algorithm is an adaptable randomized self-stabilizing protocol, which we do not formally show here. Instead, we focus on the simpler version of the 2-clock algorithm.

The 2-clock synchronization algorithm employs a shared coins algorithm, which is based on repeated invocations of a verifiable secret sharing (VSS) scheme. The VSS scheme, in turn, requires random bits to operate correctly. The random bits are supplied to the VSS algorithm by using the detector in Figure 4 at each node.

```
input: n, f, clock_i
1    Send clock_i to all nodes
2    tally_i ← |{clock_j | clock_j = clock_i}|
3    if tally_i ≥ n − f then
4    |    Return true
5    else
6    |    Return false
7    fi
```

Figure 4: $\mathcal{P}_i$ for the Byzantine clock synchronization algorithm

Next we show that the 2-clock synchronization algorithm of [3] is an adaptable randomized self-stabilizing protocol.

**Lemma 4.2.** *The 2-clock synchronization algorithm of [3] is an adaptable randomized self-stabilizing algorithm.*

*Proof.* We need to show that the detector in Figure 4 satisfies definition 4.1. First note that when the algorithm has stabilized, all the correct nodes have the same clock value. Hence, each node will receive at least $n − f − 1$ clock values identical to its own, and the detector will return *true* at all correct nodes. On the other hand, assume the system is not in a safe configuration; more specifically, let $c^i$ be a configuration in which there exist two correct nodes, $p$ and $q$, such that the clock values of $p$ and $q$ differ. Let $f'$ denote the number of Byzantine nodes in the system, such that $f' \leq f$. Let $P$ be the set of all correct nodes with the same clock value as $p$ and define $Q$ similarly with the clock value of $q$. Without loss of generality, assume that $|P| \leq |Q|$. Since $|P| + |Q| \leq n - f'$ it holds that that $|P| \leq \lfloor \frac{n-f'}{2} \rfloor$. Let $i \in P$ be a node. $i$'s tally (Figure 4, line 2), due to the non-faulty and faulty inputs, is at most $\lfloor \frac{n-f'}{2} \rfloor + f'$, and as $n > 3f \geq 3f'$, we get:

$$tally_i \leq \left\lfloor \frac{n-f'}{2} \right\rfloor + f' \leq \frac{n+f'}{2} \leq \frac{n+f}{2} < \frac{n+\frac{n}{3}}{2} = \frac{2n}{3} < n - f$$

it easily follows that $i$'s tally is at most $n - f - 1$. Therefore, $i$'s detector will return *false*, which concludes the proof. □

---

[1] Note further that [3], unlike [6], assumes the existence of a private channel connecting each pair of non-faulty nodes

The following Theorem immediately follows:

**Theorem 4.2.** *There exists a constant time, randomization adaptive Byzantine self-stabilizing clock synchronization algorithm.*

## 5 Conclusions

Memory and communications adaptive algorithms are very attractive. Taking this notion further, as random bits are expensive, using randomization only for convergence and possibly having deterministic closure is a great benefit. In a way, our approach may derandomize the closure of self-stabilizing algorithms. We presented the first fast self-stabilizing Byzantine pulse clock synchronization that uses bounded number of random bits. Further research for achieving fast clock synchronization in the settings of [6] in which non-faulty nodes do not have private channels is left for future investigation.